\documentclass[useAMS,usenatbib]{mn2e}

\title[Dissipative accretion flows around a rotating black hole]
  {Dissipative accretion flows around a rotating black hole}
\author[Santabrata Das and Sandip K. Chakrabarti]
  {Santabrata Das\thanks{sbdas@canopus.cnu.ac.kr}$^{1, 2}$
   and Sandip K. Chakrabarti\thanks{chakraba@bose.res.in} $^{3, 4}$
  \newauthor 
  \\
  $^1$ARCSEC, 98 Gunja-Dong, Gwangjin-Gu,
         Seoul 143-747, Sejong University, South Korea.\\
  $^2$Korea Astronomy and Space Science Institute
   61-1, Hwaam Dong, Yuseong-Gu, Daejeon 305 348, South Korea.\\
  $^3$ S. N. Bose National centre for Basic Sciences,
  JD-Block, Sector III, Salt Lake, Kolkata, 700 098, India.\\
  $^4$ Indian Centre for Space Physics, Chalantika 43, Garia Station Rd.,
  Kolkata 700084, India\\}

\date{\today}

\pagerange{\pageref{firstpage}--\pageref{lastpage}} \pubyear{2002}

\def\LaTeX{L\kern-.36em\raise.3ex\hbox{a}\kern-.15em
    T\kern-.1667em\lower.7ex\hbox{E}\kern-.125emX}

\usepackage{graphicx}
\input psfig.tex

\def\lsim{\mathrel{\hbox{\rlap{\hbox{\lower4pt\hbox{$\sim$}}}\hbox{$<$}}}}
\def\gsim{\mathrel{\hbox{\rlap{\hbox{\lower4pt\hbox{$\sim$}}}\hbox{$>$}}}}
\def \simeq{\lower.3ex\hbox{$\; \buildrel \sim \over - \;$}}

\begin{document}

\label{firstpage}

\maketitle

\begin{abstract}
We study the dynamical structure of a cooling dominated rotating
accretion flow around a spinning black hole. We show that non-linear
phenomena such as shock waves can be studied in terms of only three flow parameters,
namely, the specific energy (${\cal E}$), the specific angular momentum ($\lambda$) and 
the accretion rate (${\dot m}$) of the flow. 
We present all possible accretion solutions. We
find that a significant region of the parameter space in the ${\cal E}-\lambda$ plane
allows global accretion shock solutions. The effective area of the
parameter space for which the Rankine-Hugoniot shocks are possible
is maximum when the flow is dissipation free. It decreases with the increase of 
cooling effects and finally disappears when the cooling is high enough. 
We show that shock forms further away when the black hole is rotating
compared to the solution around a Schwarzschild black hole with identical flow
parameters at a large distance. However, in a normalized sense, the flow parameters 
for which the shocks form around the rotating black holes are produced shocks
closer to the hole. The location of the shock is also 
dictated by the cooling efficiency in that higher the accretion rate
(${\dot m}$), the closer is the shock location. We believe that
some of the high frequency quasi-periodic oscillations may be due to 
the flows with higher accretion rate around the rotating black holes. 

\end{abstract}


\section{Introduction}

The accretion onto black holes is thought to be an essential process  in black hole
astrophysics
since the gravitational potential energy released by the infalling
matter is directly related to the emitted power from the vicinity
of a black hole \citep{fkr02}. In the last
three decades, a number of works have reported extensive 
studies of accretion disk models around gravitating objects by the
several groups both theoretically as well as through numerical simulations
\citep{ss73,lb78,eck88,c89,nh94,yk95,c96a,ly97,ft04,nt73,pt74,hetal84a,hetal84b,g99}.
In 1980s it was realized that an
accreting matter becomes transonic around a galactic and
extragalactic black hole (BH) in order to satisfy the
inner boundary conditions at the event horizon. A fully 
general relativistic theoretical work of \citet{c96b} 
showed that multi-dimensional sub-Keplerian flows 
exhibit discontinuous shock transitions for a significant region of 
the flow parameter space spanned by the specific energy and the
specific angular momentum in a non-dissipative flow.
Recently, it has been pointed out that the shock waves formed around
a BH could be a very useful tool in explaining the observational
results. The hot, dense post-shock flow could be the natural site
for generating high energy photons \citep{cm06}. 
In particular, \citet{cm06} reported that
non-thermal electrons may be processed around the shock fronts by particle
acceleration mechanism that could produce emitted radiation up to
few MeVs by inverse Comptonization of non-thermal synchrotron
radiations. In addition, the observed quasi-periodic oscillations
(QPOs) in wide frequency range may be originated due to some type of resonance 
oscillation of the post-shock flow \citep{msc96,cam04,otm07}, or due to dynamical oscillations 
of the sub-Keplerian flows \citep{rcm97}. This oscillation is governed
by the infall time scale of matter in the post-shock region.
Since a shock may form very close to a spinning BH, it could possibly
explain the origin of high frequency QPOs also. Thus, it would be worthwhile
to study the behavior of the shock waves around a rotating BH for 
more realistic cases by addition of cooling processes inside the disk.
This is all more important since the post-shock pressure which 
holds the shocks in place, depends on the cooling processes.

So far, major progresses on the theoretical front have taken place
only when the accretion flow around the rotating BH is thin 
and when the energy dissipation processes 
are inefficient \citep{c96b}. In reality, the accreting matter should be
affected by the various cooling processes, such as bremsstrahlung and 
synchrotron cooling, most significantly the latter as 
the magnetic field is expected to be present in the disk. Some
theoretical work along this line has recently been completed \citep{d07}. However,
the effects of synchrotron cooling on the dynamical structure of the accreting flow around
a rotating black hole has not yet been studied well.
Indeed, the study of accretion flows with various energy dissipation 
processes around a compact object in the general relativistic limit is very 
complex. Recently, \citet{skcsm06} proposed a pseudo-Kerr potential which describes the 
space-time geometry around a rotating BH quite satisfactorily for the black hole spin
parameter, $a_k$, $-1 \leq a_k \leq 0.8$. This pseudo-potential, 
which is an extension of the concept laid down while writing the 
Paczynski-Wiita potential \citep{pw80} in Schwarzschild geometry
allows us to study very complex bahaviour of
dissipative accreting matter around a rotating black hole in 
a very simple and elegant way. 

In an optically thick and geometrically thin accretion flow, cooling processes
reduce the bulk energy of the flow during accretion while viscosity
not only heats the flow but also redistributes the angular momentum
in the disk. The cooling is governed by the density and temperature of the flow
while viscosity is assumed mainly due to ion collisions or its coupling with
the magnetic field. A realistic disk can have some viscosity, but 
as \citet{cd04,c96a} already pointed out, standing shock exits 
even in presence of significant viscosity. Thus, in this paper
we shall explore the properties of shock waves in presence of cooling alone.
In other words, our focus here is to deal with the accretion flows
having low angular momentum upto the marginally bound value. However,
we argue that basic conclusions regarding our idealized inviscid 
solution would be qualitatively similar to the viscous solution. 
We wish to report viscous solutions elsewhere in the future.

The suitability of the pseudo-Kerr potential that we employ has been well tested 
for both particle dynamics and fluid dynamics \citep{skcsm06,smskc06}
which allows us to use this for our present study with sufficient
confidence. We identify all possible accretion solution
topologies including the ones which may allow standing shock
transitions. In this work, we only focus on thin, non-dissipative
shock waves. We separate various regions in the parameter
space according to the solution topologies. We identify
a special region where standing shock conditions (Rankine-Hugoniot
shock conditions [RHCs], \citealt{ll59}) are not satisfied but accretion solution
continues to have three critical points. Solutions of this
kind generally exhibit oscillating shock as in \citet{rcm97} 
when Schwarzschild geometry was used. In the present paper, we ignore
bremsstrahlung cooling process since it is very inefficient compared 
to the synchrotron loss \citep{cc00}, especially in flows 
around stellar mass black holes. We also do not consider 
Compton cooling effect. These will be carried out in future.

In the next Section, the governing equations  
are presented. In \S 3 and \S 4, the accretion solutions are shown. We
discuss the shock properties in \S 5. In the final Section, we make 
concluding remarks.

\section{The Governing Equations}

In this Section, we present a set of relevant equations that governs the
dynamical structure of a stationary, thin, viscous, axisymmetric accretion
flow onto a rotating black hole. 
In addition,  we assume an ideal equation of state for accreting gas.
The equations are written on the 
equatorial plane of the accretion disk in the frame of the so-called zero angular momentum 
observers, and are given by \citep{c96a,d07},

\noindent (i) the radial momentum equation :
$$
u \frac {du}{dx}+\frac {1}{\rho}\frac {dP}{dx}+\frac{d \Phi_e}{dx}=0,  
\eqno{(1)}
$$
\noindent (ii) the mass flux conservation equation :
$$
\dot M = \Sigma u x,
\eqno{(2)}
$$
\noindent (iii) the angular momentum conservation equation :
$$
u \frac {d\lambda}{dx} + \frac{1}{\Sigma x}\frac{d}{dx}
\left( x^2 W_{x \phi}\right) = 0
\eqno(3)
$$
and
\noindent (iv) the entropy generation equation :
$$
\Sigma u T \frac {ds}{dx}
=Q^{+} - Q^{-} .
\eqno{(4)}
$$
The variables $u$, $\rho$, $P$ denote the radial velocity, 
density and isotropic pressure of the flow
respectively. In the present work, the effect of gravity is
approximated by the pseudo-Kerr potential introduced 
by \citet{skcsm06} and \citet{smskc06}. The expression of effective
pseudo-Kerr potential is given by,
 
$$
\Phi=-\frac{\Phi_2 + \sqrt{ \Phi^2_2 - 4 \Phi_1 \Phi_3}}{2 \Phi_1}
\eqno(5)
$$
where,
$
\Phi_1=\frac{\alpha^2 \lambda^2}{2x^2},
$
$
\Phi_2=-1 + \frac{\alpha^2 \omega \lambda R^2}{x^2} 
+\frac{2a_k\lambda}{R^2 x}
$
and
$
\Phi_3=1-\frac{1}{R-x_0}+\frac{2a_k\omega}{x}
+\frac{\alpha^2 \omega^2 R^4}{2x^2}.
$
Here, $x$ and $R$ represents the cylindrical and spherical radial distance
from the black hole when the black hole itself is considered to be located
at the origin of the cylindrical coordinate system. The specific angular
momentum of the flow (which is equivalent to $-u_\phi/u_t$ in a general relativistic flow, $u_\mu$
being the flow velocity components) is denoted by $\lambda$. Furthermore, 
$x_0=(0.04+0.97a_k+0.085a_k^2)/2$, $\omega=2a_k/(x^3+a^2_k x+2a^2_k)$
and $\alpha^2=(x^2-2x+a^2_k)/(x^2+a_k^2+2a^2_k/x)$.
Here, $a_k$ represents the BH rotation parameter defined as the angular
momentum of the black hole per unit mass. The subscript `e' 
stands for the quantities calculated at
the disk equatorial plan. The mass accretion rate is denoted by ${\dot M}$
and $\Sigma$ and $W_{x\phi}$ are the vertical averaged density and
the viscous stress of the flow. The terms
$s$, $T$, $Q^{+}$ and $Q^{-}$ represent the specific entropy, the local 
temperature, the energy gained and the energy lost by the flow, respectively. 
In our model, the accreting flow is assumed to be in hydrostatic 
equilibrium in the vertical direction of the flow motion. 
The half thickness $h(x)$ of the disk is then obtained by equating the vertical 
component of the gravitational force and the pressure gradient force: 
$$ 
h(x)=a \sqrt{\frac{x}{\gamma \Phi^{'}_R}},
\eqno(6) 
$$ 
where, $\Phi^{'}_R=\left(\frac{\partial \Phi}{\partial R}\right)_{z<<x}$
and $\gamma$ is the adiabatic index of the flow. 
Here, $z$ is the vertical component of the distance in the cylindrical 
co-ordinate system and  $R=\sqrt{x^2+z^2}$. The adiabatic sound speed 
is denoted by $a$ and defined as $a=\sqrt {\gamma P/\rho}$. In the
weak viscosity limit, $W_{x\phi}$ is negligible and the angular
momentum distribution remains uniform throughout the disk. 

The flow equations have been made dimensionless by considering unit of length, 
time and the mass as $GM_{BH}/c^2$,  $GM_{BH}/c^3$ and $M_{BH}$ 
respectively where, $G$ is the  gravitational constant, $M_{BH}$ is 
the mass of the black hole and $c$ is the velocity of light respectively. 
Henceforth, all the flow variables are expressed in geometrical units.

We simplify the entropy equation (Eq. 4) for an inviscid flow 
($Q^{+} \rightarrow 0$) as:
$$
\frac{u}{\gamma-1}\left[ \frac{1}{\rho} \frac{dP}{dx}
-\frac{\gamma P}{\rho^2}\frac{d\rho}{dx}\right]=\Lambda,
\eqno(4a)
$$
where, the term $\Lambda (=Q^{-}/\rho h)$ represents the energy lost per unit gram by the flow. 
In this work, we only focus on synchrotron cooling. The strength
and the morphology of the disk magnetic field are still largely
unknown. This motivated us to consider only the stochastic magnetic field 
inside the accretion disk and this magnetic field is thought to
be at the most in equipartition with the accretion plasma. We 
define a control parameter, $\beta$, as the ratio of the thermal pressure 
and the magnetic pressure of the flow which acts as a measure
of equipartition and is given by,
$$
\beta=\frac{8\pi \rho k_B T_p}{B^2 \mu m_p},
\eqno(7)
$$
where, $B$ denotes the magnetic field strength, $k_B$ is the Boltzmann 
constant, $\mu$ is the mean molecular 
weight and $m_p$ is the mass of the proton respectively. The equipartition
parameter $\beta \gsim 1$ in general, ensures that magnetic fields 
remain confined with the accreting plasma \citep{mc05}.
The synchrotron emissivity for the stochastic magnetic field is given by 
\citep{shte83,d07},
$$
\Lambda=\frac{S a^5}{u}\sqrt{ \frac{ \Phi^{'}_{R} }{x^3}},
\eqno(8)
$$
with
$$
S= 15.36 \times 10^{17}~\frac{{\dot m} 
\mu^2 e^4}{\beta m_e^3\gamma^{5/2}} \frac{1}{GM_{\sun}c^3},
\eqno(9)
$$
where, $e$ and $m_e$ represent charge and mass of an electron
respectively. Here, $\dot m$ is the accretion rate in units of
Eddington rate that regulates the efficiency of cooling. 
Following \citet{smskc06}, we use modified polytropic index 
$[n=(\gamma-1)^{-1}]$ relation as $n\rightarrow n + (0.3-0.2a_k)$
and choose $\beta = 10$ throughout the paper, until otherwise stated.

\section{Sonic point analysis}

We solve Eqs.(1-3, 4a) following the standard method of sonic point analysis
\citep{c89}. We calculate the radial velocity gradient as:
$$
\frac {d u}{dx}=\frac{N}{D},
\eqno(10)
$$
where, the numerator $N$ is given by,
$$
N = S a^5 \sqrt {\frac{\Phi^{'}_R}{x^3}}
-\frac{3 u^2 a^2}{x (\gamma -1)}
+ u^2\frac{ (\gamma+1)}{(\gamma-1)}\frac{d\Phi_e}{dx}
+ \frac{u^2 a^2}{(\gamma - 1)\Phi^{'}_R}\frac{d\Phi^{'}_R}{dx}
\eqno{(10a)}
$$
and the denominator $D$ is given by,
$$
D = \frac {2 a^2 u}{(\gamma-1)}-\frac {(\gamma+1)}
{(\gamma-1)}u^3.
\eqno{(10b)}
$$

The gradient of sound speed is obtained as:
$$
\frac{d a}{d x}=\left( \frac{a}{u} - \frac{\gamma u}{a} \right)
\frac{d u}{dx} + \frac{3 a}{2 x} 
-\frac{\gamma}{a}\frac{d\Phi_e}{dx}
-\frac{a}{2\Phi^{'}_R}\frac{d\Phi^{'}_R}{dx}.
\eqno(11)
$$
Since the matter is accreting onto the BH smoothly except at the
shock location, the radial velocity
gradient must always be real and finite. However, eq. (10b) shows that 
there may be some points where denominator ($D$) vanishes. This indicates
that the numerator ($N$) must also vanish there to keep $du/dr$ finite.
These special points where both the numerator ($N$) and denominator ($D$)
vanish simultaneously are called critical points or sonic points. 
Setting $D=0$, one can easily obtain the expression for the Mach number 
$(M = u/a)$ at the sonic point as,
$$
M(x_c) = \sqrt{\frac {2}{\gamma+1}}.
\eqno(12)
$$  

We obtain an algebraic equation for the sound speed ($a_c$) 
by using another sonic point condition $N=0$ which is given by, 

$$
F({\mathcal E}_c, \lambda_c, {\dot m})={\mathcal A}a^3_c + {\mathcal B}a_c
+ {\mathcal C}= 0 ,
\eqno(13)
$$
where,
$$
{\mathcal A}=\left[ S (\gamma-1)\sqrt{\frac{\Phi^{'}_R}{x^3}}\right]_c,
$$
$$
{\mathcal B} = \left[\left( \frac{1}{\Phi^{'}_R}\frac{d\Phi^{'}_R}{dx}
-\frac{3}{x}\right)M^2 \right]_c,
$$
and
$$
{\mathcal C} = \left[ (\gamma +1 ) M^2 \frac{d\Phi_e}{dx} \right]_c.
$$
The subscript `$c$' denotes the quantities computed at the
sonic point. We calculate sound speed at the sonic point by solving Eq. (13)
analytically \citep{as70}. In general, a dissipative accretion flow may have
multiple sonic points depending on the flow parameters. The nature of the
sonic point is dictated by the sign and the numerical value of the velocity
gradients at the sonic point. In reality $du/dr$ may possesse two real
values at the sonic point: one is for accretion branch and the other
corresponds to the wind branch. When both the values of $du/dr$ are
real and of opposite signs, the sonic point is referred to as saddle type.
A transonic flow generally passes through the saddle type sonic points
only and for a shock, the flow crosses two saddle type sonic points,
one before the shock and the other, after.
The closest one from the BH horizon is called the inner sonic point and
the furthest one is known as the outer sonic point. If the derivatives are real
and of same sign, the sonic point is nodal type. When both the derivatives 
are complex, the sonic point is of spiral type which is unphysical 
as no physical solution can pass through it. 

\section{Accretion Solution}

In order to obtain a complete accretion solution, we choose the inner sonic
point location ($x_{in}$) and the angular momentum ($\lambda$) of the flow
as input parameters \citep{d07}. From the inner sonic point, we integrate
inward up to the BH horizon and outward till the outer edge and combine
them to obtain a global transonic solution.

\subsection{Shock Solution }

\begin{figure}
\begin{center}
\includegraphics[height=9cm,width=9cm]{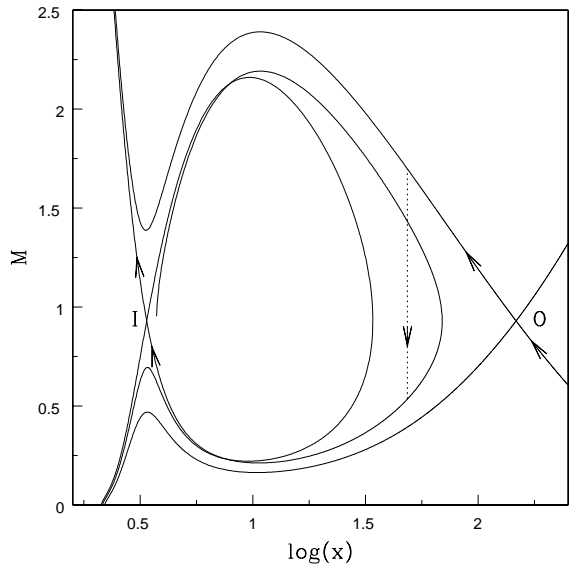}
\caption{Variation of Mach number ($M=u/a$) with logarithmic radial
distance. Flow parameters are $x_{in}=3.3753$, $\lambda =3.04$, 
${\dot m} =0.0025$ and $a_k=0.5$. Standing shock forms at $x_s = 48.57$.
}
\end{center}
\end{figure}

In Fig 1, we present a global solution with a standing shock 
around a rotating black hole. The variation of the Mach number 
with the logarithmic radial distance is plotted. The flow parameters 
are $x_{in}=3.3753$, $\lambda =3.04$, ${\dot m} = 0.0025$ and 
$a_k=0.5$ respectively.
The figure consists of two solutions. The one passing through the
outer sonic point (O) connects the black hole horizon and the
outer edge of the disk. The other passing through inner sonic point 
(I) is closed and is connected with the BH horizon only. Arrows indicate the 
direction of the flow towards the BH. Matter starts accreting 
from the outer edge of the disk with a negligible velocity. As 
the flow accretes towards the BH, the flow gains its radial velocity 
due to the attraction of the strong gravity. The flow becomes super-sonic 
after crossing the outer sonic point (O) and continues to 
accrete towards the BH. The RHCs \citep{ll59} 
in turn allows the flow variables to make a discontinuous 
jump in the sub-sonic branch. This is indicated by a dotted vertical line and 
commonly known as the $shock$ transition. In the post-shock 
region, the flow momentarily slows down and subsequently picks up the radial
velocity and enters into the BH supersonically after crossing the
inner sonic point (I). In this particular case, the shock 
conditions are satisfied at $x_s = 48.57$. The entropy generated at the shock
is eventually advected towards the BH to allow the flow to pass 
through the inner sonic points. In this work, the shocks are considered to be
thin and non-dissipative.

\subsection{Parameter space for multiple sonic points}

\begin{figure}
\begin{center}
\includegraphics[height=9cm,width=9cm]{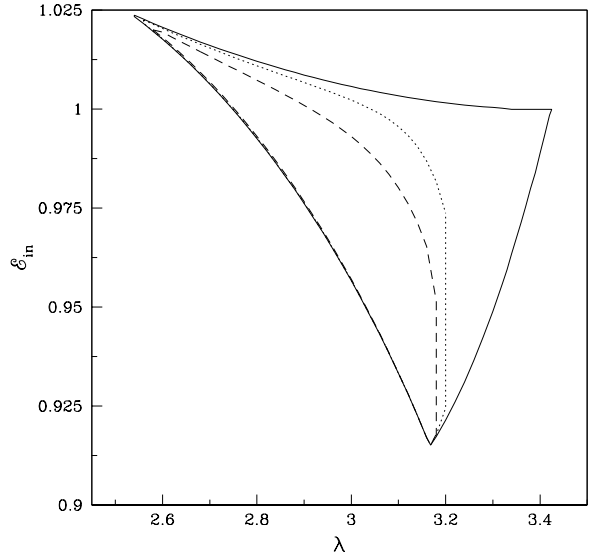}
\caption{Parameter space for multiple sonic points. The effective region
bounded by the solid curve are for cooling free accretion flow. Regions under
dotted and dashed curves are obtained for ${\dot m } = 0.0025$
and $0.0125$ respectively. Effective area of parameter space reduces 
with the increase of accretion rate $({\dot m})$ as it enhances the cooling
efficiency.
}
\end{center}
\end{figure}

So far, we have seen that the shock wave connects two solution branches---one
passing through the outer sonic point and the other passing through the 
inner sonic point. In particular, the flow with multiple saddle type 
sonic points may undergo shock transitions. More importantly, for shock 
formation in dissipative accretion flows, the solution passing through
the inner sonic point has to be spiraling in \citep{d07}. 
Therefore, it would be useful to study the parameter space for accretion flows 
having spiral in solution passing through the inner sonic point.  
In Fig. 2, we show the classifications of parameter space as a function 
of accretion rate ($\dot m$) in the ${\cal E}_{in}-\lambda$ plane. 
Here, ${\cal E}_{in}$ denotes the energy of the flow at the inner sonic 
point ($x_{in}$). The BH rotation parameter is chosen as $a_k=0.5$. The region
bounded by the solid curve is obtained for non-dissipative
accretion flows. The regions under the dotted and dashed curve is obtained
for higher accretion rates ${\dot m} = 0.0025$ and $0.0125$ respectively. 
As accretion rate (${\dot m}$) is increased, the parameter space for multiple
sonic points shrinks. This 
indicates that nature of sonic points changes (from saddle to spiral) for 
flow with identical input parameters for increasing ${\dot m}$. 

\section{Shock properties}
\subsection{Shock Dynamics}

\begin{figure}
\begin{center}
\includegraphics[height=9cm,width=9cm]{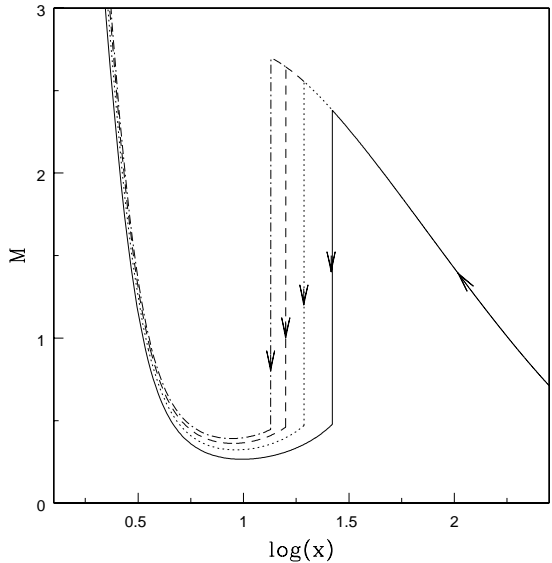}
\caption{Mach number variation with logarithmic radial distances.
Flows are injected sub-sonically from the outer edge $x_{inj}=300$
with identical energy ${\cal E}_{inj}=1.003163$ and angular momentum
$\lambda= 3.0$. Different accretion rates (${\dot m}$) are used.
Solid curve represents a solution including the shock wave ($x_s= 26.03$)
for cooling free accretion flow. Other solutions are for
[${\dot m}$, $x_s$]= [0.0125, 19.38] (dotted), [0.025, 15.85] (dashed)
and [0.0375, 13.45] (dot-dashed). As ${\dot m}$ is increased, shock front
precedes towards BH.
}
\end{center}
\end{figure}

In Fig. 3, we present the variation of shock locations with the accretion rate (${\dot m}$).
Logarithmic radial distance is varied along the horizontal axis and  
Mach number is plotted along vertical axis. The vertical lines represent
the shock locations. Matter with identical outer boundary conditions is
injected sub-sonically from the outer edge of the disk $x_{inj}=300$ on to
a rotating BH with rotation parameter $a_k = 0.5$.
The local energy of the flow at $x_{inj}$ is ${\cal E}_{inj}\equiv 
{\cal E}(x_{inj})=1.003163$ (including the rest mass) and the angular 
momentum is $\lambda = 3.0$. The sub-sonic flow crosses the outer sonic 
point to become super-sonic and makes a shock transition to the 
sub-sonic branch. The solid vertical lines represent the shock 
location ($x_s = 26.03$) for non-dissipative flow. As
cooling is incorporated, shock front moves forward. In the post-shock
region, cooling is more effective compared to the pre-shock flow as 
the density as well as the temperature are very high in this 
region due to compression. Cooling reduces the post-shock pressure
causing the shock front to move inward to maintain pressure balance
across it. For higher ${\dot m}$, shock front
moves further inward. The dotted, dashed and dot-dashed vertical lines 
represent the shock locations $x_s = 19.38, 15.85$ and $13.45$ for ${\dot m}
=0.0125, 0.025$ and $0.0375$ respectively.

\begin{figure}
\begin{center}
\includegraphics[height=9cm,width=9cm]{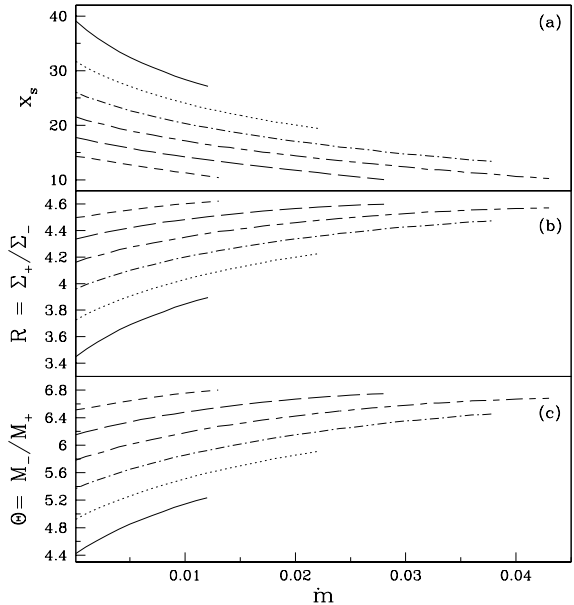}
\caption{(a) Variation of the shock location with accretion rate (${\dot m}$).
Flows are injected with the same energy and angular momentum from the outer
edge. Small dashed, big dashed, small-big dashed, dot-dashed, dotted 
and solid curves are drawn for angular momentum $\lambda = 2.94, 2.96, 2.98,
3.0, 3.02$ and $3.04$ respectively. (b) Variation of compression ratio
($R=\Sigma_{+}/\Sigma_{-}$) with accretion rate for the same set of
parameters as in (a). (c) Variation of shock strength ($\Theta = M_{-}/M_{+}$)
with the accretion rate for the same set of parameters as in (a). Subscripts
``+" and ``-" denote quantities before and after the shock. }
\end{center}
\end{figure}

In Fig. 4a, we show the variation of the shock location as a function of
accretion rate (${\dot m}$) for a set of angular momentum ($\lambda$). 
In this particular figure, the flow is injected from the outer edge of the 
disk ($x_{inj} = 300$). The BH rotation parameter is chosen as $a_k= 0.5$.
The angular momentum of the flow is varied from 
$\lambda = 2.94$ (small dashed) to $3.04$ (solid) with an interval 
$\Delta \lambda = 0.02$. At the injection point, the corresponding 
local energies of the flow (from bottom to top) 
are ${\cal E}_{inj}= 1.003174$ (small dashed), $1.003170$ (big dashed), 
$1.003167$ (small-big dashed), $1.003163$ (dot-dashed), $1.003160$ (dotted), 
and $1.003156$ (solid) respectively. For a given accretion rate (${\dot m}$), 
the shock forms further out for flows with higher angular momentum. Here,
the larger angular momentum increases the centrifugal pressure which
pushes the shock front outside. Conversely, for a given angular
momentum, the shock location decreases with the increase of the accretion rate
as cooling reduces the post-shock thermal pressure. 
Figure 4a shows that the standing shocks are formed for a wide 
range of accretion rate (${\dot m}$). In each angular momentum, the
standing shocks disappear beyond a critical value of accretion rate
(${\dot m}_c$) as the RHCs are not satisfied here. 
Non-steady shocks may still exit, but an investigation of such phenomena 
is beyond the scope of the present work. As $\lambda$ increases, the critical 
accretion rate first increases, becomes maximum at some $\lambda$ 
($=2.96$, in this particular case) and then decreases. This clearly indicates
that the parameter space for the standing shock shrinks in both the lower and higher
angular momentum sides with the increase of the accretion rate.
 
One of the important components in accretion disk physics
is to study the density profile of matter since the cooling 
efficiency as well as the emitted radiation directly depends 
on it. We compute the compression ratio $R$ defined as the
ratio of vertically averaged post-shock to pre-shock density
and plot it in Fig. 4b as a function of accretion rate (${\dot m}$) 
for the same set of flow parameters as in Fig. 4a. For a given angular 
momentum ($\lambda$), the compression ratio increases monotonically with 
higher ${\dot m}$. 
As ${\dot m}$ increases, post-shock flow becomes more compressed to provide
required pressure for holding the shock.
In addition, for a given ${\dot m}$, higher angular momentum flow feels 
less compression in the post-shock region as centrifugal 
pressure resists the flow to accrete. Note that, for each angular 
momentum, there is a cut-off at a critical accretion rate limit as 
standing shock conditions are not satisfied there. 

It is useful to study the another shock property called shock strength $\Theta$
(defined as the ratio of pre-shock to post-shock Mach number of the flow)
as it is directly related to the temperature jump at the shock.  
In Fig. 4c, we show the variation of shock strength as a function of 
accretion rate (${\dot m}$) for flows with identical input parameters
as in Fig. 4a. For a given angular momentum $(\lambda)$, the strength of the
shock is the weakest in the dissipation-free limit and it becomes stronger as 
accretion rate (${\dot m}$) is increased. Thus, a higher cooling causes 
the post-shock flow to be hotter and radiations emitted from this region are
expected to be harder. A similar result is reported by \citet{mc05}.
This clearly indicates that the observed spectra of the BH would 
strongly depend on the cooling.

\begin{figure}
\begin{center}
\includegraphics[height=9cm,width=9cm]{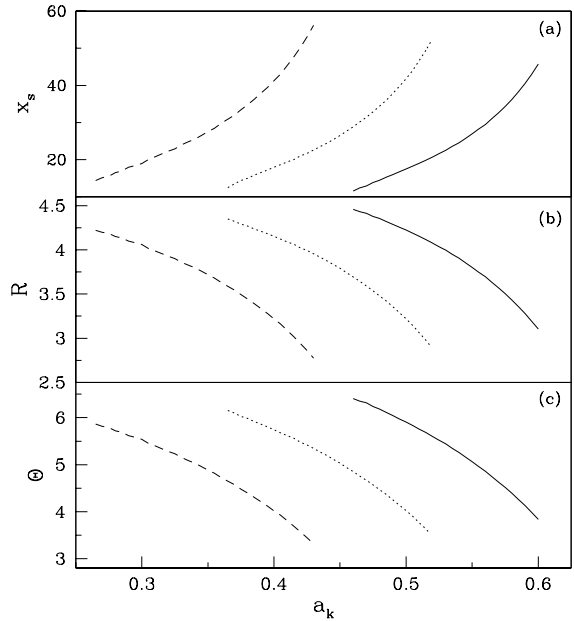}
\caption{Variation of (a) shock location, (b) compression ratio
and (c) shock strength with BH rotation parameter $a_k$. See text for
more details. 
}
\end{center}
\end{figure}

An important part of understanding a cooling dominated accreting flow
around a rotating BH is to study the shock properties as a function
of BH rotation parameter $a_k$. In Fig. 5a, we plot the variation of
shock location as a function of $a_k$. In this particular Figure, we 
inject matter from the outer edge $x_{inj}= 200$ and the accretion rate
is considered to be ${\dot m} = 0.0025$. Solid, dotted and dashed curves are
obtained for flows with angular momentum $\lambda = 2.96, 3.05$ and $3.14$
respectively. The corresponding energies at the injection point are
${\cal E}_{inj} = 1.003456, 1.003442$ and  $1.003422$ respectively. 
Notice that, for a given $\lambda$, shocks form for a particular range 
of $a_k$. As $a_k$ increased, shock recedes from the BH horizon. Moreover, 
shocks exist around a weakly rotating BH when the flow angular momentum 
is relatively higher and {\it vice versa}. This phenomenon is directly
related to the 
spin-orbit coupling term in the Kerr geometry. In fact, since both
the marginally bound and the marginally stable angular momenta
(as well as their difference) go down when the Kerr parameter is 
increased, the relevant parameter region when the shocks form also goes down
as $a_k$ is increased. In general, however, the shock location is generally 
small for higher $a_k$, as statistically the flow with a smaller angular
momentum (and therefore, the lesser centrifugal force) is accreted in 
a rapidly spinning black hole. Thus, for instance if we compared the shock
locations having $\lambda=\lambda_{ms}$ for all cases, the shock location for
a rotating black hole would be closer for higher $a_k$.

In Fig. 5b, we show the variation of the compression ratio as a function of
BH rotation parameter $a_k$ for the flows with input parameters same
as Fig 5a. The solid, dotted and dashed curves are obtained for $\lambda
= 2.96, 3.05$ and $3.14$ respectively. The compression ratio $R$ decreases 
with the increase of $a_k$ for flow with identical $\lambda$. In Fig. 5c, 
we plot the variation of shock strength $\Theta$ with $a_k$ for flow 
with input parameters as in 
Fig. 5a. We obtain a similar variation of $\Theta$ with $a_k$ as in Fig. 5b. 

\subsection{Parameter Space for Shock Formation}

\begin{figure}
\begin{center}
\includegraphics[height=9cm,width=9cm]{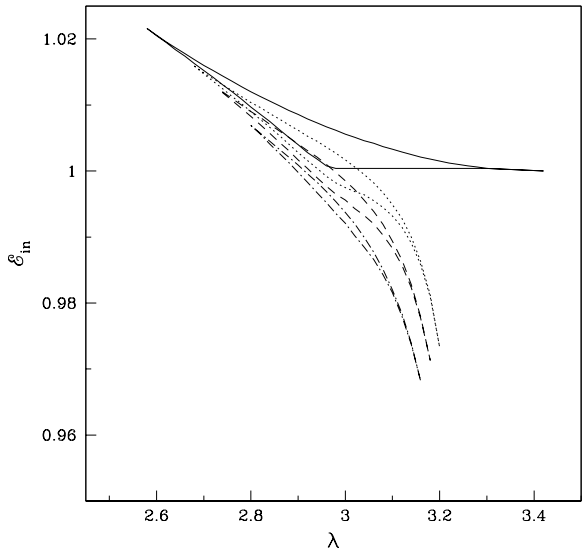}
\caption{Variation of the effective region of parameter space which forms 
standing shocks as a function of the accretion rate (${\dot m}$). See the
text for more details.
}
\end{center}
\end{figure}

In Fig. 6, we identify the region of the parameter space that allows
the formation of the standing shocks. The BH rotation parameter 
is considered to be $a_k = 0.5$. 
The region bounded by the solid curve is obtained for 
non-dissipative accretion flow. As accretion rate is enhanced, the effective
region of parameter space for standing shocks shrinks in both the 
lower and higher angular momentum sides. Due to the cooling effect, the flow
loses its energy as it accretes and therefore, the parameter space is shifted
to the lower energy domain for higher cooling. The regions
under dotted, dashed and dot-dashed curves are obtained for accretion
rates ${\dot m} = 0.0025, 0.005$ and $0.01$ respectively. It is clear that the
standing shocks do not exist beyond a critical accretion rate when the
synchrotron cooling is present.

\begin{figure}
\begin{center}
\includegraphics[height=8cm,width=8cm]{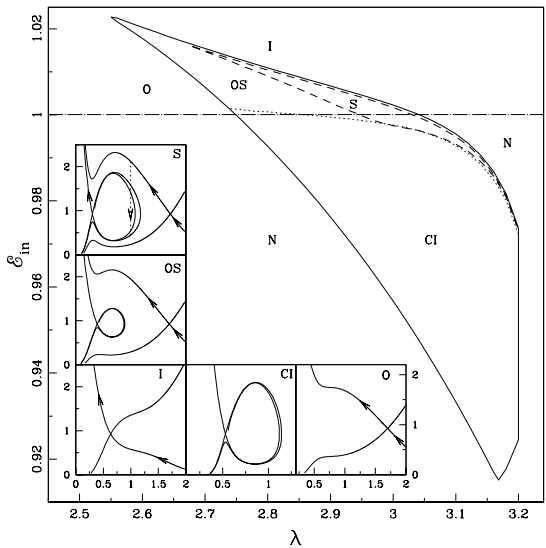}
\caption{Classification of parameter space according to the various
solution topologies of BH accretion solution. See text for details.
}
\end{center}
\end{figure}

In Fig. 7, we classified the entire parameter space spanned by 
(${\cal E}_{in}, \lambda$) according to the nature of solution topologies. 
As an example, we consider $a_k = 0.5$ and ${\dot m} = 0.0025$.
We separate the parameter 
space into six regions marked by S, OS, O, I, CI and N. The dot-dashed
line represents the rest mass energy of the flow. At the bottom left of the
parameter space, we plot solution topologies in the small boxes. In each
box, Mach number of the flow is plotted against the logarithmic radial distance.
Each of these solutions are marked and drawn using the parameters from 
the corresponding region of the parameter space. The direction of the 
accreting flow is indicated by the arrow. (The solutions without
arrows are relevant for winds, discussions on which are
beyond the scope of this paper.) The solutions from the regions marked
`S' and 'OS' have two X type sonic points and the entropy at the inner 
sonic point is higher than that at the outer sonic point. 
Flows from `S' suffer a standing shock 
transition as RHCs are satisfied. However, a solution from the region `OS' 
does not pass through the standing shock as RHCs are not satisfied here.
A flow with parameters from this region is unstable and causes periodic
variation of emergent radiation from the inner part of the disk as it
tries to make a shock transition but fails to do so. This is known from the
numerical simulations of non-dissipative flows \cite{rcm97} 
and we anticipate that a similar behaviour would be seen in 
this case as well. The solutions from the region `I'
possess only the inner sonic point and the accreting solutions 
straight away pass through it before entering into 
the BH. A solution from the region `O' has only
one outer sonic point. The solution from region `CI' has two sonic points---
one 'X' type and other 'spiral' type. Solutions of this kind does not extend to the
outer edge of the disk to produce a complete global solution and 
therefore, becomes unstable. It has been pointed out by
\cite{c96a} that inclusion of viscosity should open up the
topology to allow the flow to reach a larger distance
to join with a sub-sonic Keplerian flow. 
The region marked `N' is the forbidden region 
for a transonic flow solution.

\section{Concluding Remarks}

In this paper, we have studied the properties of cooling dominated accretion
flow around a rotating black hole by solving a set of equations that regulate
the dynamical structure of the flow. A special consideration is given to
synchrotron cooling that strongly affects the disk properties as well as
the emitted spectrum and luminosity that are observed.
 
We obtain the global accretion solutions with and without the shocks in terms
of a few flow parameters, namely, energy, angular momentum, 
black hole rotation parameter and the accretion rate, which effectively
acts as the cooling parameter. We find that the accreting matter experiences a
centrifugal force which acts as a barrier, inducing a shock formation. We
show that the global shocked accretion solution can be obtained for a significant
region of the parameter space even when the cooling is significant. Using a conventional
accretion disk model we expect the accretion to take place when the angular momentum
is close to the marginally stable value. Our calculation shows that
the region is actually broader, in terms of both the angular momentum and energy.

The discussion regarding the nature of the sonic point has been 
reported in many occasions \citep{cd04,dc04}. However, a detailed analysis
was not presented before for a cooling dominated flow around a
rotating black hole. Our present work suggests that a large region
of the parameter space provides a stable saddle type sonic point. In Fig. 2,
we demonstrated that the parameter space for the stable saddle type sonic point
is gradually reduced with the increase of cooling efficiency.      

We show that the standing shocks form closer to a spinning black hole as the
accretion rate is enhanced. At the post shock region, the density and the 
temperature is relatively high compared to the
pre-shock flow and thus cooling is more efficient there. For a higher
cooling, the post-shock matter cools faster reducing the
thermal pressure drastically. This forces the shocks to move inward to
maintain pressure balance across them. 

One of the aims of the present work was to study the effect of
black hole rotation parameter on the dynamical structure of cooling
dominated global solutions. We find that for flows with identical
outer boundary condition ({\it e. g.,} same energy and angular momentum
at the outer edge) shock recedes from the black hole horizon with the
increase of black hole rotation parameter ($a_k$). However if we 
choose the relevant angular momentum for each case, such as the marginally stable angular momentum
the shock location moves in with the increase of $a_k$. The
range of $a_k$ for which the stationary shocks are formed is restricted
for a flow of given angular momentum. Shocks are possible around a 
rapidly rotating black hole when the flow angular momentum is 
relatively low. Since that produces a very low centrifugal
pressure, the shock can form very close to the black hole
for a rapidly spinning black hole. 

We identify the region of parameter space for the formation of
a standing shock. We find that the effective region of the
parameter space for the stationary shock shrinks when
the accretion rate is enhanced. This suggests 
that the possibility of shock formation decreases 
for higher accretion rate. In addition, we also 
separate a region where the Rankine-Hugoniot relation is not
satisfied. In the context of invidcid flows, it has been observed that the
flow parameters from such a region give rise to oscillating shocks 
\citep{rcm97}. The reason is that the higher entropy
at the inner sonic point forces the flow to pass through it by generating
extra entropy at the shock. But since RHCs are not satisfied the 
shock can not settle itself at a given location. Thus the cause of oscillation is
sufficiently generic and we suspect that exactly the same thing will happen in the
present case. Most importantly, since rotating black holes may have shocks
very close to the horizon, the frequencies of such oscillations 
are expected to be higher.

Our present findings suggest that shocks, standing or oscillating,
do form around the spinning black holes and it may be an essential
ingredient since shocks could successfully explain the observed 
stationary \citep{cm06} as well as time dependent behaviour of
the radiations from the black hole candidates \citep{cam04,otm07}. We demonstrated
that the shocks form closer to the black hole as cooling is increased. 
This will enhance the QPO frequency as it is proportional to the infall 
time scale \citep{cm00,msc96}
and thus the QPO frequency may vary in a wide range starting
from mHz to KHz depending on the accretion rate. This understanding
also generally agrees with the observational results. Recent
reporting of the outbursts of GRO 1655-40 showed a clear evidence of 
the QPO frequencies increasing monotonically
from about $90$mHz to $17$Hz \citep{cd05} in a matter of $~ 15 $ days
and this could be easily fitted using the shock propagating 
at a constant velocity (Chakrabarti et al. 2005).

The formalism presented here does not include outflows/jets which may
be generated from the inner part of the disk as a result of deflection
of inflowing matter due to excess thermal pressure at the shock front
\citep{dcnc01,cd07,dc08}.
Since the outflows/jets are ejected evacuating the inner part of the
disk, it will necessarily reduce the post-shock pressure and therefore,
shock front has to move in to retain the pressure balance. This
suggests that the result should be affected if the accretion-ejection
mechanism is considered together. We plan to consider this study in
a future work and it will be reported elsewhere. 

We have approximated the effect of general relativity using the pseudo-Kerr
gravitational potential. This pseudo-Kerr potential has been 
successfully tested to retain most, if not all, of the salient
features of the flows in a Kerr metric. The use of this approach 
allows us to find out the non-linear
shock solutions in a curved space-time geometry in a simpler way. 
We believe that our basic results would be qualitatively the same with fully 
general relativistic calculations, especially for $a_k <0.8$ for which 
the pseudo-Kerr potential was found to be satisfactory. 

\section*{Acknowledgments}
SD was supported by KOSEF through Astrophysical Research Center
for the Structure and Evolution of the Cosmos(ARCSEC). SKC thanks
a visit to Abdus Salam International Centre for Theoretical Physics
where part of this work was completed.

\end{document}